# Mobile recommender systems: Identifying the major concepts

**Elias Pimenidis**
Department of Computer Science and Creative Technologies, University of the West of England, Bristol, BS16 1QY, United Kingdom

**Nikolaos Polatidis\* and Haralambos Mouratidis**
School of Computing, Engineering and Mathematics, University of Brighton, Brighton, BN2 4GJ, United Kingdom

\*Corresponding author, N.Polatidis@Brighton.ac.uk

**Abstract**
This paper identifies the factors that have an impact on mobile recommender systems. Recommender systems have become a technology that has been widely used by various online applications in situations where there is an information overload problem. Numerous applications such as e-Commerce, video platforms and social networks provide personalized recommendations to their users and this has improved the user experience and vendor revenues. The development of recommender systems has been focused mostly on the proposal of new algorithms that provide more accurate recommendations. However, the use of mobile devices and the rapid growth of the internet and networking infrastructure has brought the necessity of using mobile recommender systems. The links between web and mobile recommender systems are described along with how the recommendations in mobile environments can be improved. This work is focused on identifying the links between web and mobile recommender systems and to provide solid future directions that aim to lead in a more integrated mobile recommendation domain.

## 1. Introduction

Recommender systems research is becoming increasingly important in e-Commerce environments. Since their emergence popular recommendation systems exist and include Netflix, Amazon, YouTube, MovieLens and Epinions among others. Furthermore, well-known recommendation libraries include Apache Mahout, Lenskit and Recsys. More specifically, a recommender system uses a certain algorithm or techniques in order to suggest items or services of interest to users, thus aiming to solve in a way the information overload problem found in the World Wide Web (Bobadilla, Ortega, Hernando, & Gutiérrez, 2013; Jannach, Zanker, Felfernig, & Friedrich, 2010; Lu, Wu, Mao, Wang, & Zhang, 2015). The recommendations include movies, books, songs, software, tourism related material, jokes, general products and even people to people in social networks. The recommender system adapts to each user in a personalized way to provide recommendations, which is usually done through previous preferences, current interests or a combination of methods. The provided recommendations are suggestions of items or services that could be of potential interest to the user. For example, in a movie recommender system scenario the list of recommended items would include a set of movies that are expected to be of an interest to the user.

Recommender systems, depending on the method they employ can be classified in one of the following categories (Bobadilla et al., 2013; Jannach et al., 2010; Shi, Larson, & Hanjalic, 2014; Su & Khoshgoftaar, 2009):

**Collaborative filtering.** These recommenders suggest items to users that other users with a similar rating history have liked in the past.

**Content-based.** These recommenders suggest items to users that are similar to the items that user has liked in the past.

**Knowledge-based.** These recommenders suggest items to users based either on inferences about the preferences of users or by utilizing specific domain knowledge.

**Hybrid.** These recommenders are based on a combination of two or more algorithms.

However, privacy is an important aspect in various e-Commerce environments and everyone should have a right to it. In mobile recommendation scenarios apart from the recommendation method employed other important aspects include the utilization of context variables and the protection of user privacy (Gavalas, Konstantopoulos, Mastakas, & Pantziou, 2014; Polatidis, Georgiadis, Pimenidis, & Stiakakis, 2017; Rodríguez-Hernández & Ilarri, 2016). Thus, if a recommended item is relevant this is due to the recommendation method and the context. Therefore, mobile recommender systems should make suggestions to users based on a recommendation method, the available contextual information and in addition user privacy needs to be preserved. A typical overview of mobile recommender systems is shown in figure 1.

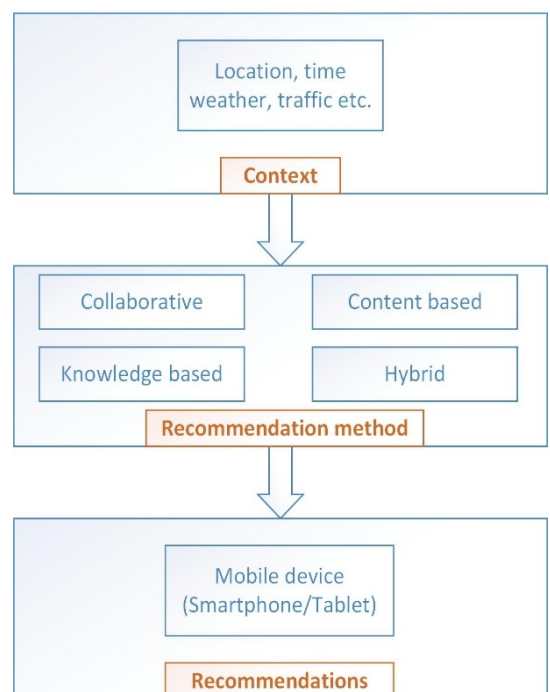

**Figure 1.** Recommendation process in mobile environments

In this paper, our goal is to (a) analyze the major concepts that affect mobile recommender systems, (b) explain how these can be improved and (c) provide suggestions for future research. The rest of the paper is organized as follows: Section 2 contains background work, section 3 explains the findings, section 4 contains the conclusions and presents future research directions.

## 2. Background

### 2.1 Related work

Recommender systems are usually based on a 2-dimensional setting to provide recommendations to users. This means that typical contextual information such as the time, the weather and the location of the user requesting the recommendations are typically ignored. Furthermore, users mostly use a mobile device such as a smartphone or tablet for most of their tasks and perform those while on the move. Mobile recommender systems are based on a recommendation algorithm and contextual information to provide recommendations of items or services to users of mobile devices. Thus, a set of components that can be utilized to facilitate the process of developing mobile recommender systems is delivered.

Several problems have been identified and solved to a certain extend. For example, in (Lathia, 2015) an overview of mobile recommender systems is given with the outcome that a mobile recommendation domain does not exist and that every mobile recommender system is developed with a specific task in mind. This argument is also present in other works in the literature such as (Jannach et al., 2010; Q. Liu, Ma, Chen, & Xiong, 2013; Rodríguez-Hernández & Ilarri, 2016). In addition, certain works such as Hybreed (Hussein, Linder, Gaulke, & Ziegler, 2014) develop on this concept. In Hybreed the authors propose a framework for the automatic delivery of context-aware hybrid recommender systems. Also, the term UbiCars has been proposed by (Mettouris & Papadopoulos, 2014). In this survey work, the characteristics of ubiquitous computing and recommender systems are described in detail and the authors explain what ubiquitous recommender systems are. Another, relevant work has been proposed by (Rodríguez-Hernández & Ilarri, 2016). In this work, the process of developing pull-based mobile recommender systems is explained in detail and a generic mobile recommendation architecture is proposed and evaluated. Furthermore, one of the major issues identified is the absence of good relevant datasets in the literature and that most datasets are domain specific. Consequently, an important problem identified it as the lack of a good quality datasets for the purpose and the research is based on specific mobile domains. These include the development of mobile recommender systems for the recommendations of generic commerce items like books, movies, music and photos or the recommendations of points of interest (POIs) such as restaurants and tourist attractions. Thus, the remaining part of this chapter concentrates on the coverage of related works found both in the general commerce and in the tourism domains respectively.

Mobile recommender systems for the recommendation of books, photos and music have been proposed in the literature. For example, CoMeR (Yu et al., 2006) provides the recommendation of different types of media to its users using a context-aware approach. Also, in (Lemos, Carmo, Viana, & Andrade, 2012) the prototype version of a photo recommender system can be found. It is a mobile recommender system for photos which utilizes current contextual data in combination with information found in the photos. Another interesting work is found in (Baltrunas et al., 2011) with the name of InCarMusic. A context-aware recommender system is used to provide music recommendations to the passengers of a car. In addition, another similar work for the recommendation of music depending on the daily activities of a person is provided by (Wang, Rosenblum, & Wang, 2012). This mobile recommender is based on a probabilistic model to propose songs to users of mobile devices depending on their current activity. Another example of a mobile recommender system can be found in the mobile news domain that is based on the current context and the format of the recommended news (Sotsenko, Jansen, & Milrad, 2014). Mobile recommender systems have been used for movie show times. RecomMetz (Colombo-Mendoza, Valencia-García, Rodríguez-González, Alor-Hernández, & Samper-Zapater, 2015) is based on a knowledge based recommender and contextual information to recommender movie show times to users of mobile devices. Another interesting M-commerce recommender system for the recommendation of mobile applications based on collaborative filtering and context has been proposed by (J. Lin et al., 2011). Motivate is yet another mobile recommender for the recommendation of personalized activities for the user who wants to maintain a healthy lifestyle (Y. Lin, Jessurun, De Vries, & Timmermans, 2011). Furthermore, in mobile recommender different data sources can be utilized to provide personalized recommendations to users. These include the approach proposed by (Zhu et al., 2014) were mobile logs are used and the method proposed by (D. R. Liu & Liou, 2011) that utilizes data from multiple mobile channels. Thus, it is noticeable that different recommendation methods and data sources can be used along with contextual information for servicing users.

SMARTMUSEUM is a mobile recommender system proposed by (Ruotsalo et al., 2013). The aim of this recommender is to provide recommendations for heritage related items in indoor and outdoor scenarios. To achieve this, different type of contextual information is utilized such as the location and time. Another mobile recommender system based on collaborative filtering and context data is iTravel (Yang & Hwang, 2013). This is a peer-to-peer recommender for the recommendation of travel attractions. On the other hand, Turist is a mobile recommender system for recommending cultural and leisure activities once the user is at the destination (Batet, Moreno, Sánchez, Isern, & Valls, 2012). Additionally, mobile recommender systems can be used to provide pushed recommendations without explicit requests from the user. The recommender recommends item or services when the context is appropriate for those. Such a recommender system has been proposed by (Woerndl, Huebner, Bader, & Gallego-Vico, 2011). Furthermore, context-aware collaborative filtering has been used by (Huang & Gartner, 2012) for the recommendation of mobile guides. In this work, the authors recommend similar POIs to users with similar interests in similar contexts. A mobile recommender system based on banking data history has been proposed (Gallego & Huecas, 2012). The mobile application recommends places that users have visited before and paid with their credit card. A modified collaborative filtering approach that uses several contextual variables has been proposed to provide recommendations using a decision tree and a set of rules (Kim, Ahn, & Jeong, 2010). A mobile recommender system for guides that is based on collaborative filtering and contextual information has been proposed by (Gavalas & Kenteris, 2011).

## 2.2 Factors affecting mobile recommender systems

Three factors exist that can affect mobile recommender systems and their ability to provide accurate personalized recommendations. These include the context, the recommendation method and privacy (Polatidis & Georgiadis, 2014).

### 2.2.1 Context

Context is used by mobile recommender systems to provide accurate personalized recommendations to users who are constantly moving. Different context types can be used according to the scenario and could include amongst others: The location, time and the weather. Location-based services are based on contextual data and information can be collected either implicitly or explicitly (Adomavicius & Tuzhilin, 2015; Ricci, 2010). Furthermore, context-awareness in mobile computing is a paradigm where applications can discover contextual information using the sensors of the mobile device and an application can automatically adjust its behavior according to the context. In contrast, desktop computers can use context variables for their applications but not adapt in changing environments well. In addition, mobile context information might come from other environmental such as traffic or peers that are in proximity. Although these variables can be entered in desktop computing as well, this has to be done in a manual way (Musumba & Nyongesa, 2013; Polatidis, Georgiadis, Pimenidis, & Stiakakis, 2017; Sarwat et al., 2015).

The application of the context variables in the recommendation process can be made in three different ways (Adomavicius & Tuzhilin, 2015):

**Contextual pre-filtering.** This is a method where context is used to filter out any data that are not relevant and then the recommendation method is applied.

**Contextual post-filtering.** This is a method where the recommendations method is initially used to derive recommendations and then the context is applied to sort out any irrelevant data.

**Contextual modeling.** This is a method where the recommendation function is designed in a way that the context is utilized within.

### 2.2.2 Recommendation method
A recommendation method is used to suggest items or services to users and is a necessary part of mobile recommender systems. The most widely known recommendation methods are (Bobadilla et al., 2013; Lu et al., 2015): Collaborative filtering. This is a method that utilizes similar rating history between users to recommend items or services. This works by asking users to submit numerical ratings and then a function is used to search between user ratings for similar users and provide the recommendations (Shi et al., 2014). Content-based filtering is a method that is based on keywords supplied by users and uses them to match item descriptions (Konstan & Riedl, 2012). Knowledge based filtering is a method that is based on inferences and domain knowledge to make recommendations. Finally, hybrid methods use a combination of two or more methods to provide recommendations (Jannach et al., 2010).

### 2.2.3 Privacy
Mobile recommender systems offer the user the possibility of receiving personalized recommendations in constantly moving environments. However, this possibility comes with a privacy cost, since user data might be processed in unexpected ways by service provides. Thus, the user attitude towards the mobile recommendation process might be negative (Mettouris & Papadopoulos, 2014). At present most privacy protection techniques rely on location protection, while steps have been made towards the protection of different context data (Polatidis, Georgiadis, Pimenidis, & Stiakakis, 2017). Privacy is considered very important and should be addressed properly, thus making mobile recommender systems more usable and protect user data.

## 2.3 Motivating scenario

In this section, we describe a scenario that motivates towards the use of mobile recommender systems. This scenario shows the benefits that a user gains if a mobile recommender system is used. The example that follows is from a fictional user Bob who uses a fictional mobile application MobiApp, which can be installed in smartphones or tablets. This application recommends movies to users according to previous rating history and available context parameters.

### 2.3.1 Example scenario

Alice is at her home with her friend Bob on a Saturday evening, while outside it is raining. They are relaxing until to the point where Bob brings up the idea of watching a movie. Since, they can't decide what to watch, Bob, offers to use MobiApp in his smartphone to assist him in finding a movie to watch with his friend Alice. Then, he opens the application and selects the option of receiving recommendations to his screen. MobiApp then uses collaborative filtering to find relevant recommendations and applies contextual information (such as location, company) to sort irrelevant movies out. Figure 2 gives a high-level overview of how MobiApp works.

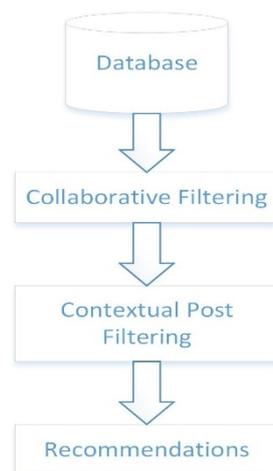

**Figure 2.** MobiApp recommendation process

MobiApp maintains a database where registered users of the app can submit their ratings. The recommender then communicates with the database and then applies collaborative filtering to come up with a list of top-N recommendations. At the next step, the contextual information is sent from the mobile device to the server so the irrelevant movies according to the current context are not recommended. At this moment the server has received various private information from the user, which needs to be protected using appropriate measures (Kato et al., 2012; Kido, Yanagisawa, & Satoh, 2005; Polatidis, Georgiadis, Pimenidis, & Stiakakis, 2017). Finally, at the end of the movie Bob is asked by MobiApp to rate the movie they just watched with Alice. An appropriate user interface pops up in the screen offering different numerical values for Bob to choose from. The user can now freely submit his rating for the movie.

In summary the steps made by Bob to produce a list of top-N recommendations in his smartphone are as follows:

1. Opens MobiApp.
2. Selects the option 'receive recommendations'.
3. MobiApp applies collaborative filtering based on Bob's rating history.
4. MobiApp applies contextual specified parameters (e.g. location and company).
5. The list of top-N recommendations is displayed.

It is important to understand that the list of the recommendations is initially created from Bob's rating history but the order of the top-N recommendation and the decision if a recommendation will be included in the list of the displayed recommendations is made after the contextual parameters are taken into consideration. For example, if Bob is alone at home and his mood is down a different list of recommendations will be provided compared when he is located at Alice's house, the weather is bad, but both feel great. In (Košir, A., Odic, A., Kunaver, M., Tkalcic, M., & Tasic, 2011) there is a detailed explanation of contextual variables that have been identified as useful in context-aware movie recommendation environments. The description of the 12 contextual variables is shown in table 1 and these variables have been used with collaborative filtering in their LDOS-CoMoDa movie recommendation dataset. A mobile recommender system either for movies or other products or services should be specifically developed with the application domain in mind and permit the user to either use contextual variables automatically or enter any of the information manually. Furthermore, the statistical description of the dataset is shown in table 2. The LDOS-CoMoDa dataset is based on user ratings from 1 to 5 from 95 users on 961 items. Assuming the one of the users is Bob then some of the ratings belong to him. These ratings among with similar ratings from other users in the database will be used from a collaborative filtering algorithm to identify a list of temporary top-N recommendations which will change after the request for the inclusion of contextual data. Finally, note that for this scenario collaborative filtering has been used there are other types of recommenders such as content-based, knowledge based and hybrid approaches can be used in the recommendation process (Bobadilla et al., 2013). In addition to that, the contextual filtering can be done either before the recommendation method is used, after the recommendation method is used or using recommendation methods with integrated contextual filtering (Adomavicius & Tuzhilin, 2015).

| **Table 1.** Description of Context Variables of LDOS-CoMoDa dataset | **Table 1.** Description of Context Variables of LDOS-CoMoDa dataset | **Table 1.** Description of Context Variables of LDOS-CoMoDa dataset |
|---|---|---|
| Time | 1 to 4 | 1=morning, 2=afternoon, 3=evening, 4=night |
| Daytype | 1 to 3 | 1=working, 2=weekend, 3=holiday |
| Season | 1 to 4 | 1=spring, 2=summer, 3=autumn, 4=winter |
| Location | 1 to 3 | 1=home, 2=public, 3=friend's house |
| Weather | 1 to 5 | 1=sunny, 2=rainy, 3=stormy, 4=snowy, 5=cloudy |
| Social | 1 to 7 | 1=alone, 2=partner, 3=friends, 4=colleagues, 5=parents, 6=public, 7=family |
| endEmo | 1 to 7 | 1=sad, 2=happy, 3=scared, 4=surprised, |

|  |  | 5=angry, 6=disgusted, 7=neutral |
| --- | --- | --- |
| dominantEmo | 1 to 7 | 1=sad, 2=happy, 3=scared, 4=surprised, 5=angry, 6=disgusted, 7=neutral |
| Mood | 1 to 3 | 1=positive, 2=neutral, 3=negative |
| Physical | 1 to 2 | 1=healthy, 2=ill |
| Decision | 1 to 2 | 1=By user, 2=By other |
| Interaction | 1 to 2 | 1=first, 2=number of interactions, after first |

**Table 1.** Description of Context Variables of LDOS-CoMoDa dataset

| Description | Value |
| --- | --- |
| Users | 95 |
| Items | 961 |
| Ratings | 1665 |
| Average age of users | 27 |
| Countries | 6 |
| Cities | 18 |
| Maximum submitted ratings from one user | 220 |
| Minimum submitted ratings from one user | 1 |

**Table 2.** Statistical description of LDOS-CoMoDa dataset

## 3. Findings

### 3.1 Links between mobile and web-based recommender systems

Mobile recommender systems develop on top of e-Commerce recommender systems with a specific mobile recommendation scenario in mind (Lathia, 2015; Polatidis & Georgiadis, 2014; Polatidis, Georgiadis, Pimenidis, & Stiakakis, 2017; Ricci, 2010; Rodríguez-Hernández & Ilarri, 2016). Thus web-based technologies need to be utilized and adapted according to the scenario. Web-based recommender systems are based mostly on a recommendation method and on a privacy preservation method such as:

- Collaborative filtering.
- A method that preserves user privacy while user ratings are submitted and used by collaborative filtering methods.

In addition, mobile environments need to:

- Utilize context parameters to provide personalized recommendations based on variables such as the location, the weather and time of day among others.

Recommendations provided in mobile devices typically use collaborative filtering as the recommendation algorithm, a privacy preservation method that perturbs user submitted ratings before these are submitted to the server, utilizes context parameters and are applied according to a specific recommendation scenario in mind according to the requirements.

Thus, we identify three research problems that need to be taken into consideration while developing mobile recommender systems.

1. Collaborative filtering
2. Privacy-preserving collaborative filtering
3. Context privacy

**3.1.1 How can the quality of collaborative filtering recommendations be improved?**
Collaborative filtering is widely used in web and mobile environments to provide suggestions of items or services to users, which include the recommendations of movies, books, general products or users to users (Jannach et al., 2010; Polatidis & Georgiadis, 2016; Shi et al., 2014). Therefore, users can benefit from finding relevant items without the burden of manually searching and companies could grow extra profit from higher sales of items. Additionally, the computational needs for the company are reduced, since the users normally will not bother making several search attempts to find relevant items. Recommender systems and collaborative filtering are valuable tools for both a user and a company and if an algorithm can produce quality results is beneficial for both sides. In the literature there are several methods that improve the quality of recommender system output, when compared to traditional methods, by providing more relevant top-N recommendations to the user (Gan & Jiang, 2013; Polatidis & Georgiadis, 2016). The quality of relevant top-N recommendations can be measured in terms of precision and recall, which is beyond the scope of this paper. Relevant information about evaluating recommender systems can be found in (Bellogin, Castells, & Cantador, 2011; Herlocker, Konstan, Terveen, & Riedl, 2004). Different recommendation methods offer different outputs according the dataset and how they work. In terms of precision and recall the higher the value the higher the quality of the recommendations.

Thus, in H1 we hypothesize that:

*H1 (a): The quality of collaborative filtering recommendations in mobile environments should be as high as possible.*
*H1 (b): Users of mobile devices want to find the most relevant recommendations in the top of their list when compared to users of web-based systems.*

This hypothesis can be validated according to the application domain and the scenario. When a mobile recommender system is developed, then it should use real data and be used by real users in the context that it has been developed for. Then, the feedback received by the users can be used to validate the results. There are studies available that have been used in such situations (Anacleto, Figueiredo, Almeida, & Novais, 2014; Baltrunas et al., 2011; Baltrunas, Ludwig, Peer, & Ricci, 2012).

**3.1.2 How can user privacy be protected when collaborative filtering systems are used?**
Privacy-preserving collaborative filtering methods can be used to protect user privacy. Different protection methods can be found in the literature (Boutet, A., Frey, D., Guerraoui, R., Jégou, A., & Kermarrec, 2016; Polatidis, Georgiadis, Pimenidis, & Mouratidis, 2017;

Songjie Gong, 2011; Tada, Kikuchi, & Puntheeranurak, 2010; Yakut & Polat, 2012). These methods add a level of noise to the data, thus providing a level of protection which comes with a decrease in terms of the quality of the top-N recommendations. Any of these methods can be used to perturb data and then a recommendation method can use the perturbed data to provide usable recommendations.

Thus, in H2 we hypothesize that:

*H2: Users of mobile devices want a layer of protection like the needs of web-based systems users.*

This hypothesis has been based on the idea that privacy-preserving collaborative filtering in web based recommender systems is important. Thus, we assume that users of mobile recommender systems that utilized collaborative filtering will want to preserve their privacy as well. To validate this hypothesis a user based survey should be conducted. This survey should present the facts of privacy and how personal data can be used by third, most of the time unauthorized, parties.

### 3.1.3 How can user privacy be protected when context parameters are utilized?
Privacy is an important issue in recommender systems and most techniques are either based on collaborative filtering rating privacy or location privacy. With users moving constantly and by using different contextual variables to receive personalized recommendations, all these data are communicated with services providers (Gamecho et al., 2015; Polatidis, Georgiadis, Pimenidis, & Stiakakis, 2017). Although, in the literature some work towards the protection of context data has been done, more work towards this direction will be necessary (Polatidis, Georgiadis, Pimenidis, & Stiakakis, 2017).

Thus, in H3 we hypothesize that:

*H3 (a): Context is a very important aspect in mobile recommender systems.*
*H3 (b): The use of context variables is necessary to receive recommendations of high quality in mobile recommender systems.*
*H3 (c): It is vital to protect user privacy, while context variables are used.*

This hypothesis has been proved to an extend in specific studies that involved the use of mobile recommender systems in various domains including tourism and commerce (Anacleto et al., 2014; Fang et al., 2012; Yang & Hwang, 2013). To validate this hypothesis, software developers or researchers should develop a mobile recommender system for a specific application domain and ask real users to use it and receive feedback on the outcome with and without the use of context-variables. Furthermore, it should be made clear to the users of such a system what the privacy concerns regarding context variables are. The protection of context aware recommender system variables in mobile recommender systems is explained in detail in (Polatidis, Georgiadis, Pimenidis, & Stiakakis, 2017).

### 4. Conclusions and directions for future research
This work examined the concepts behind the development of mobile recommender systems and concludes that there exists a gap between recommender systems and mobile computing that needs to be filled. Mobile computing and recommender systems should be more tightly integrated, thus resulting in the research field of mobile recommender systems. Now, various web-based recommendations technologies (such as recommendation and privacy algorithms) are modified to be used in mobile devices. Recommender systems in mobile environments carry most of the characteristics found in e-Commerce recommenders. However, it is noted that the use of algorithm selection or privacy preserving technologies should be selected based on the domain and the scenario. Furthermore, any method combination could fit well together but every system has its strengths and limitations. In addition, not all methods mentioned here might be appropriate for the scenario in mind. Provided that the research components are combined and tested in various mobile recommendation scenarios then more useful conclusions could be derived. Furthermore, stronger conclusions can be derived with the use of the components in scenarios where real users are engaged, and observational evaluation techniques used to monitor the users using the mobile recommender system. Mobile recommender systems build on top of web-based recommender systems and with a specific scenario in mind. For this purpose, web-based techniques are used and in the literature most work is directed towards the development of specific mobile recommender systems or towards the improvement of accuracy or privacy techniques. Improvements in these directions are useful for recommender systems but not enough to bridge the gap between mobile computing and recommender systems. Research needs to take place in various parts and how these can be combined to offer integrated solutions. Suggestions for future research are listed below.

**A framework for the delivery of mobile recommender systems.** One future research direction is the delivery of a complete framework for mobile recommender systems, with an aim of bridging the gap between mobile computing and recommender systems. Relevant work in mobile recommender systems points towards a direction where an integrated framework needs to exist. Thus, a part of future work could concentrate on the proposal of a framework for developing mobile recommender systems. The framework will utilize different recommendation methods, use them according to the current scenario and provide different options for privacy preserving recommendations depending on the recommendation algorithm and context settings used. However, this research direction includes the evaluation of the system using real world data, well established metrics and by engaging with real users in every stage of the process. Furthermore, each method of such the framework should be evaluated separately from the others and as an integrated whole at the end.

**Resolve the complexity of context data.** A second future research direction will have to deal with the complexity of context data. Towards this direction both structured and unstructured data can be collected using different sources. The form of these data tends to be both complex and large, thus it will be difficult to process them. It will be necessary to develop appropriate technologies to deal with the storing, search and retrieval of contextual information and as a result more information will be available, which will result in more accurate recommendations.

**Personalized advertisements.** A third research direction relates to the concept that current mobile recommender systems are developed with specific requirements in mind (e.g. for specific tourism tasks). However, with the availability of different data from different sources the possibility of understanding users better and provide more personalized advertisements or marketing of certain products or services that are more relevant to both the user and its current scenario could be delivered.

**Explanations.** A fourth research direction should be directed towards the explanation of the recommendations. Since, explanations are very important in recommender systems and with the special characteristic of mobile devices different ways should be found to provide satisfactory explanations to users.

**Privacy.** A fifth research direction will be towards context privacy protection. Although initial steps have been made more extensive work will need to take place for protecting the privacy of users utilizing enormous amounts of context data while the speed and the accuracy of the recommender system remains reasonable.

**Spam.** A sixth research direction will have to be towards spam protection. A big challenge for recommender systems is the protection from shilling attacks. Positive or negative ratings are inserted deliberately to promote certain products and while certain measures exist to prevent those, these do not apply to contextual data. Nowadays, with the popularity of mobile devices increasing and with mobile recommender systems available for different domains, research will need to take place to provide systems that will protect users of context-aware mobile recommender systems from possible attackers.